\documentclass[pdfusetitle,aps,reprint,prd,twocolumn,superscriptaddress,nofootinbib]{revtex4-1}
\pdfoutput=1

\usepackage{amssymb}
\usepackage{amsmath}
\usepackage{epsfig}
\usepackage{breakurl}
\usepackage{float}
\usepackage{multirow}
\usepackage{array}
\usepackage{bm}
\usepackage{color}
\usepackage{tikz}
\usepackage[utf8]{inputenc}
\usepackage{braket}
\usepackage{graphicx}
\usepackage{footmisc}
\usepackage{tensor}
\usepackage[normalem]{ulem}
\usepackage{dblfloatfix}
\usepackage[T1]{fontenc}
\usepackage{array}
\usepackage{booktabs}
\usepackage{multirow}
\usepackage{amstext}
\usepackage{multirow}



\usepackage{babel}

\usepackage[force]{feynmp-auto}

\xdefinecolor{mylinkcolor}{rgb}{0,0,0.7}
\usepackage[
	bookmarksnumbered, bookmarksopen, bookmarksopenlevel=2,
	breaklinks=true,
	colorlinks=true, filecolor=mylinkcolor, citecolor=mylinkcolor,
	linkcolor=mylinkcolor, urlcolor=mylinkcolor, menucolor=mylinkcolor,
]{hyperref}
\usepackage{bookmark}

\usepackage{cleveref}
\crefformat{section}{\S#2#1#3} 
\crefformat{subsection}{\S#2#1#3}
\crefformat{subsubsection}{\S#2#1#3}

\usepackage{colortbl}
\definecolor{blue2}{cmyk}{1, 0.1, 0.1, 0}

\definecolor{pyBlue}{RGB}{31, 119, 180}
\definecolor{pyRed}{RGB}{214, 39, 40}
\definecolor{pyGreen}{RGB}{44, 160, 44}
\definecolor{pyBlue2}{RGB}{0, 111, 237}
\definecolor{pyRed2}{RGB}{224, 52, 36}

\definecolor{summersky}{cmyk}{0.71,0.33,0,0.5}
\definecolor{flamingo}{cmyk}{0,0.51,0.71,0.5}
\definecolor{rp}{cmyk}{0.2, 1, 0.6, 0}
\definecolor{pacificblue}{cmyk}{0.95,0.3,0, 0.5}
\definecolor{gray60}{cmyk}{0.4,0.4,0,0.8}

\renewcommand{\vec}[1]{\mathbf{#1}}

\usetikzlibrary{decorations.markings}
\usepackage{cleveref}
\crefformat{section}{\S#2#1#3}
\crefformat{subsection}{\S#2#1#3}
\crefformat{subsubsection}{\S#2#1#3}

\makeatletter
\def\simgt{\mathrel{\lower$\frac{5}{2}$pt\vbox{\lineskip=0pt\baselineskip=0pt
           \hbox{$>$}\hbox{$\sim$}}}}
\def\simlt{\mathrel{\lower$\frac{5}{2}$pt\vbox{\lineskip=0pt\baselineskip=0pt
           \hbox{$<$}\hbox{$\sim$}}}}

\makeatother

\def\spa#1.#2{\left\langle#1\,#2\right\rangle}
\def\spb#1.#2{\left[#1\,#2\right]}
\def\sand#1.#2.#3{%
\left\langle#1{\vphantom1}\right|{#2}\left|#3\right]}%
\def\sandmp#1.#2.#3{%
\left\langle#1{\vphantom1}\right|{#2}\left|#3\right]}%
\def\sandpm#1.#2.#3{%
\left[#1{\vphantom1}\right|{#2}\left|#3\right\rangle}%
\def\sandmm#1.#2.#3{%
\left\langle#1{\vphantom1}\right|{#2}\left|#3\right\rangle}%
\def\sandpp#1.#2.#3{%
\left[#1{\vphantom1}\right|{#2}\left|#3\right]}%

\renewcommand{\imath}{\mathrm{i}}

\newcommand{\be}{\begin{equation}}
\newcommand{\ee}{\end{equation}}

\def\S{{\mathbb S}}

\allowdisplaybreaks

\begin{document}

\title{A New Type of Saddle in the Euclidean IKKT Matrix Model\\ and Its Emergent Geometry}

\author{Henry Liao}
\email{henryliao.physics@gmail.com}
\affiliation{Department of Physics and Center for Theoretical Physics, National Taiwan University, Taipei 10617, Taiwan}

\author{Reishi Maeta}
\email{maeta-reishi@hiroshima-u.ac.jp}
\affiliation{Department of Physics and Center for Theoretical Physics, National Taiwan University, Taipei 10617, Taiwan}
\affiliation{Graduate School of Advanced Science and Engineering, Hiroshima University, Higashi-Hiroshima
739-8526, Japan}
\affiliation{Physics Department, McGill University, Montreal, QC H3A 2T8, Canada}

\begin{abstract}
We study the equation of motion of the Euclidean IKKT matrix model, and realize a new type of classical saddle that only exists in $N\rightarrow\infty$ limit.
Under the assumption that the matrices are the generators of $\mathfrak{so}(n,m)$, we identify a unique solution, that is, $\mathfrak{so}(1,3)$.
Even though it has $6$ generators and thus $6$ non-zero matrices, they are not independent due to the $2$ Casimir constraints in $\mathfrak{so}(1,3)$.
Exploiting the Lie-algebraic structure and the Casimir constraints, we derive a four-dimensional space that a test scalar propagates on. 
The associated metric possesses $\mathrm{SU}(2)$ isometry, which is closely related to the Taub–NUT/Bolt geometry and, more broadly, to black hole physics. 
\end{abstract}

\maketitle

\section{Introduction}

As a promising proposal toward non-perturbative formulation of type IIB superstring theory, the IKKT matrix model \cite{Ishibashi:1996xs, Aoki:1998bq} is expected to describe non-perturbative objects such as a $D3$-brane that we live on.
In fact, this zero-dimensional matrix model suggests a picture that four-dimensional spacetime can be emerged dynamically~\cite{Kim:2011cr, Anagnostopoulos:2022dak, Nishimura:2019qal, Brandenberger:2024ddi}. 
In addition, its low energy effective theory is closely related to general relativity and can be unitary~\cite{Ho:2025htr}.

There are other attempts to understand the IKKT matrix model. 
A holographic dictionary between type IIB supergravity and the IKKT matrix model has been established~\cite{Ciceri:2025maa, Ciceri:2025wpb}. 
For the so-called polarised IKKT matrix model~\cite{Bonelli:2002mb}, in which special deformations are added while preserving maximal supersymmetry, there are similar discussions~\cite{Hartnoll:2024csr, Hartnoll:2025ecj, Komatsu:2024bop, Komatsu:2024ydh}.
For other types of deformations, various studies have shown the emergence of
four-dimensional near-realistic spacetime, which has FLRW metric~\cite{Steinacker:2017vqw, Sperling:2019xar, Battista:2022hqn} or even chiral fermions~\cite{Nishimura:2013moa, Chatzistavrakidis:2011gs}.

Among these various developments, from the viewpoint of describing the real world in which we live, the original IKKT matrix model without mass deformations is the most suitable for detailed analysis. This is because the IKKT matrix model is derived either (1) by the \textit{large-$N$ reduction} of ten-dimensional $\mathcal{N}=1$ supersymmetric Yang--Mills theory~\cite{Eguchi:1982nm, Parisi:1982gp, GROSS1982440}, or (2) by \textit{matrix regularization} of the Green--Schwarz action of type IIB superstring theory~\cite{Ishibashi:1996xs} (for type IIA superstring theory, see also~\cite{deWit:1988wri, Banks:1996vh}), and in both derivations no mass term is present.

However, for the massless theory, only a limited number of classical solutions are currently known~\cite{Ishibashi:1996xs,Manta:2025tcl,Chatzistavrakidis:2011su,Shibusa:2003dg}, let alone solutions describing four-dimensional spacetime.
Motivated by this situation, we try to find a new saddle of the original IKKT matrix model which is defined in Euclidean signature.
More concretely, we examined whether configurations of the form $A_\mu = M_{ab}$, constructed from the generators $M_{ab}$ of $\mathfrak{so}(n,m)$, satisfy the equations of motion~\eqref{eq:ikkt_eom}. As a result, we found that among these cases, $\mathfrak{so}(1,3)$ is the only solution.
Furthermore, we analysed which
geometry is emergent from such a saddle.
It turns out that the resulting geometry describes a four-dimensional Euclidean spacetime with $\mathrm{SU}(2)$ isometry, which bears a strong resemblance to the Taub--Nut/Bolt geometry~\cite{Taub:1950ez, Newman:1963yy}. So far no example in which black hole physics emerges from the IKKT matrix model has been known, and the $\mathfrak{so}(1,3)$ solution may constitute the first such example.

In the beginning of Sec.~\ref{sec:classical_saddles}, we briefly mention our setup, and show the action and the equation of motion in the IKKT matrix model.
Then, we mention a standard statement related to the equation of motion in the Euclidean IKKT matrix model, see Sec.~\ref{sec:standard_statement_loophole}.
In there, we point out a loophole to the statement occurs when the matrices are infinite-dimensional.
This motivates us to consider, in Sec.~\ref{sec:sonm}, an ansatz that the matrices are infinite dimensional.
More specifically, they are generators of $\mathfrak{so}(n,m)$ with $n,m\geq 1$, which only have infinite dimensional unitary representations.\footnote{The $\mathfrak{so}(n,m)$ is different from the spacetime Lorentz symmetry in general.}
We find that, under our ansatz, $\mathfrak{so}(1,3)$ is the only possibility.
This is a complete new class of solution, see \ref{sec:classification}.
To analyse its emergent geometry, we consider a test scalar matrix to find the effective metric in semi-classical limit.
This procedure aligns with that in \cite{Steinacker:2007dq}, and we provide the detail in Sec.~\ref{sec:emergent_geometry}.
Then, we conclude in \ref{sec:conclusion_and_outlook}.

\section{Classical saddles of Euclidean Bosonic IKKT matrix model}\label{sec:classical_saddles}

In the Euclidean IKKT matrix model, the action of the bosonic sector reads
\begin{align}\label{eq:bosonic_action}
    S_{\text{IKKT,B}}=-\frac{1}{4}\mathrm{tr}\left(\delta_{\mu\rho}\delta_{\nu\sigma}\left[A^\mu,A^\nu\right]\left[A^\rho,A^\sigma\right]\right).
\end{align}
The equation of motion for $A^\mu$ is given by
\begin{align}\label{eq:ikkt_eom}
    \delta_{\mu\rho}\left[A^\mu,\left[A^{\rho},A^{\sigma}\right]\right] = 0,
\end{align}
which can also be regarded as equation of motion of the full IKKT matrix model but with the assumption that the fermionic matrices set to $0$.
Note that during the derivation, we have assumed that the cyclic property of trace is always valid, and this is assumed throughout this paper.

\subsection{A Standard Statement and Its Loophole}
\label{sec:standard_statement_loophole}

A standard statement of the solutions for the Euclidean equation of motion \eqref{eq:ikkt_eom} goes as follows.
Starting with \eqref{eq:ikkt_eom}, we multiply from left an $\delta_{\nu\sigma}A^\nu$ and take trace which yields
\begin{equation}
\begin{split}
    0
    &=\mathrm{tr}\left(\delta_{\nu\sigma}A^\nu\delta_{\mu\rho}\left[A^\mu,\left[A^{\rho},A^{\sigma}\right]\right]\right)\\
    &=-\mathrm{tr}\left(\delta_{\mu\rho}\delta_{\nu\sigma}\left[A^\mu,A^\nu\right]\left[A^\rho,A^\sigma\right]\right)\\
    &=\mathrm{tr}\left(\left(i\left[A_\mu,A_\nu\right]\right)\left(i\left[A_\mu,A_\nu\right]\right)\right)\label{eq:trace_com}.
\end{split}
\end{equation}
Since $A_\mu$'s are Hermitian matrices, the combination $i\left[A_\mu,A_\nu\right]$ is also Hermitian;
for each pair of $\mu,\nu$, $i\left[A_\mu,A_\nu\right]$ has real eigenvalues and thus $\mathrm{tr}\left(i\left[A_\mu,A_\nu\right]\right)^2$ is non-negative.
Hence, the last line of \eqref{eq:trace_com} can only vanish if all eigenvalues of $i\left[A_\mu,A_\nu\right]$ vanish for all $\mu,\nu$, meaning $A^\mu$'s commute.

Even though this statement sounds solid, it implicitly assumes $A^\mu$'s are finite in size.
In fact, it is clear that Heisenberg algebra, which is defined by $\left[X,Y\right]=Z$, $\left[X,Z\right]=0$ and $\left[Y,Z\right]=0$, is a solution of~\eqref{eq:ikkt_eom}.
Moreover, some nilpotent and solvable Lie algebras are solutions as well~\cite{Chatzistavrakidis:2011su}.
The loophole stems from that the argument intrinsically assumes that the trace operation is only over finite terms, that is, having finite unitary representations.
However, the exceptional cases mentioned above do not have finite dimensional unitary representations, which protects them from the argument.
Hence, to look for other types of classical saddles, one must consider cases where $A^\mu$'s are infinite dimensional.

\subsection{Classification of the IKKT saddles}\label{sec:classification}

Before discussing our solution, we make a brief detour on a classification of the saddles that can occur in \eqref{eq:ikkt_eom}.

With simple observation, \eqref{eq:ikkt_eom} contains three operations: (a) the inner commutator, (b) the outer commutator and (c) the summation over spacetime indices.
To satisfy the equation of motion \eqref{eq:ikkt_eom}, we can have the solution to be vanished at either one of the steps, meaning three different categories of solutions.

The solutions mentioned previously fall either into (a) or (b).
In particular, the case where $\left[X,Y\right]=0$ falls into (a), while the cases of Heisenberg, solvable and nilpotent Lie algebras are contained in (b).
There is no known third class solution so far, until now.
In particular, the solution we found below is the first known example of the class (c) in IKKT matrix model.

\subsection{$\mathfrak{so}(n,m)$ solution}
\label{sec:sonm}

In general, it is difficult to control matrices of infinite size.
To make this possible, we consider Lie algebra ansatz for $A^\mu$, which allows us to operate commutator algebraically without specifying the representation in prior.
The size of matrices is then determined by specifying the unitary representation of the chosen Lie algebra ansatz.
In particular, we are seeking for matrices of infinite size that solves \eqref{eq:ikkt_eom}.
A convenient and physical choise is to consider the case of the Lorentz algebra with signature $\left(n,m\right)$.
Since when $n,m\geq 1$, the corresponding unitary representations are infinite dimensional \cite{tung1985group}, we expect that, if solutions exist, $n,m\geq 1$ is required.

The convention we used to define the Lorentz algebra is as follows
\begin{equation}\label{eq:lorentz_algebra_com}
\begin{split}
    &\left[M_{ab},M_{cd}\right] \\
    &= -i\left( h_{ad}M_{bc}+h_{bc}M_{ad}-h_{ac}M_{bd}-h_{bd}M_{ac} \right)
\end{split}
\end{equation}
where $a, b$ $=$ $1,\dots,d$ with $d=n+m$ and $h_{ab}$ is the signature matrix with $n$ and $m$ numbers of $+1$ and $-1$, respectively.
Next, we assume that $A_{\mu}$'s have one-to-one correspondence to the generators $M_{ab}$'s without any further linear combinations.
That is, the dimension of spacetime is aligned with that of the algebra, which is $\frac{d\left(d-1\right)}{2}$, and the explicit ansatz is written as
\begin{align}
    \left(A_1,A_2,\dots,A_{\frac{d(d-1)}{2}}\right)=\left(M_{01},M_{02},\dots,M_{(d-1)(d)}\right).
\end{align}

In terms of the generators of the Lorentz algebra, the equation of motion~\eqref{eq:ikkt_eom} now reads
\begin{align}\label{eq:eom_of_M}
    \sum_{a,b}\left[M_{ab}\left[M_{ab},M_{ij}\right]\right]=0.
\end{align}
After double applications of \eqref{eq:lorentz_algebra_com}, we obtain
\begin{align}\label{eq:h}
    2\left( h_{ii}+h_{jj} \right)\left( \text{tr}\left(h\right) - h_{ii} - h_{jj} \right)M_{ij}=0
\end{align}
for all $i,j=1,\dots,d$.
The equality holds when one of the following two situations is true: 
\begin{enumerate}
    \item $h_{ii}=-h_{jj}$
    \item $\mathrm{tr}\left(h\right)=h_{ii}+h_{jj}$.
\end{enumerate}
The generators of the Lorentz algebra are responsible for either boosts or rotations.

For the case of boosts, we have one timelike index and one spacelike index for $i,j$, namely, $h_{ii}=-h_{jj}$.\footnote{An index $i$ is called timelike or spacelike, if $h_{ii}=+1$ or $h_{ii}=-1$, respectively. This terminology is not related to spacetime.}
Hence, the boost generators automatically satisfied the equation of motion.

On the other hand, the rotation generators do not satisfy the first case and are restricted by the second one.
Since $i,j$ can be both timelike or both spacelike for rotation generators, we have two equations
\begin{align}
    \mathrm{tr}(h)&=+2\\
    \mathrm{tr}(h)&=-2,\label{eq:spacelike}
\end{align}
where the first one is the equation for rotation in timelike directions, while the second one is that in spacelike directions.
Obviously, we cannot satisfy them simultaneously.
Hence, the only possibility to remove the contradiction is to have only one timelike or one spacelike index, which means we only have to satisfy one of them.
Below, we consider the case of a single timelike index, meaning we set $\mathfrak{so}(n,m)$ to $\mathfrak{so}(1,m)$.
That means we only need to satisfy \eqref{eq:spacelike}, which reads
\begin{align}
    1-m=-2
\end{align}
implying $m=3$.
Hence, we get $\mathfrak{so}(1,3)$ generators as our solution.
Through similar calculation with the case of a single spacelike index, we get $\mathfrak{so}(3,1)$.
Overall, within our ansatz with $\mathfrak{so}(n,m)$, the only solution is $\mathfrak{so}(1,3)$!

Unlike the case of $\mathfrak{so}(3)$ that $S^2$ is natural to be interpreted as the emerged geometry, it is not intuitive to argue the emergent geometry of this $\mathfrak{so}(1,3)$ solution.
In fact, we cannot even argue that the emerged geometry possess $SO(1,3)$ isometry, since the ambient metric is Euclidean.
Hence, in the following, we proceed to discuss how a metric can be emerged from this solution in the Euclidean IKKT matrix model.

\section{Emergent geometry from the $\mathfrak{so}(1,3)$ matrix background}
\label{sec:emergent_geometry}

In this section, we aim to extract the geometry emerged from $\mathfrak{so}(1,3)$ matrix background in semi-classical limit.
We first describe the underlying four-dimensional coadjoint orbit and its Poisson structure, and then introduce a test scalar matrix whose semi-classical equation of motion allows us to read off the effective metric. 
This demonstrates how a nontrivial curved geometry emerges from the $\mathfrak{so}(1,3)$ matrix configuration.

\subsection{Coadjoint orbit of $\mathfrak{so}(1,3)$ and Poisson structure}

We consider six embedding coordinates $a^\mu$ ($\mu=1,\dots,6$) which realize the
$\mathfrak{so}(1,3)$ algebra (in semi-classical limit) and satisfy two independent Casimir constraints. 
With the real parameters $C_1$ and $C_2$ specifying the values of the two Casimir invariants of $\mathfrak{so}(1,3)$, the Casimir functions read
\begin{equation}
\begin{split}\label{eq:casimirs}
    C_1 &= \sum_{i=1}^3 x_i^2 - y_i^2, \\
    C_2 &= \sum_{i=1}^3 x_i y_i,
\end{split}
\end{equation}
where $(x_1,x_2,x_3,y_1,y_2,y_3)=(a^1,a^2,a^3,a^4,a^5,a^6)$.
Fixing $(C_1,C_2)$ defines a four-dimensional coadjoint orbit, which we denote by $\mathcal{M}_4$.

We introduce local coordinates $s^a$ ($a=1,\dots,4$) on $\mathcal{M}_4$ and write $a^\mu = a^\mu(s)$. 
A convenient choice is
\begin{equation}
    s^a = (r,\theta,\varphi,\chi),
\end{equation}
where $r \in [r_*,\infty)$ with $r_*$ being the lower bound that will be explicit soon below, $(\theta,\varphi)$ parameterize an $S^2$, and
$\chi$ parametrizes an $S^1$. 
An explicit chart is given by
\begin{equation}\label{eq:chart}
\begin{split}
    \vec{x} &= \sqrt{C_1 + r^2}\,\hat{u},\\
    \vec{y} &= \frac{C_2}{\sqrt{C_1 + r^2}}\,\hat{u}
              + \sqrt{r^2 - \frac{C_2^2}{C_1 + r^2}}\,\hat{w},
\end{split}
\end{equation}
where
\begin{equation}
\begin{split}
    \hat{u} &= (\sin\theta\cos\varphi,\ \sin\theta\sin\varphi,\ \cos\theta),\\
    \hat{w} &= \cos\chi\,\hat{e}_1 + \sin\chi\,\hat{e}_2,
\end{split}
\end{equation}
and
\begin{equation}
\begin{split}
    \hat{e}_1 &= (\cos\theta\cos\varphi,\ \cos\theta\sin\varphi,\ -\sin\theta),\\
    \hat{e}_2 &= (-\sin\varphi,\ \cos\varphi,\ 0).
\end{split}
\end{equation}
In this chart, the boundary is at $\sqrt{r^2 - \frac{C_2^2}{C_1 + r^2}}=0$, which gives 
\begin{align}\label{eq:r_*}
    r_*=\sqrt{\frac{-C_1+\sqrt{{C_1}^2+4{C_2}^2}}{2}}>0.
\end{align}
By construction, $\hat{u},\hat{e}_1,\hat{e}_2$ form an orthonormal frame on $S^2$, and $\hat{w}$ is a unit vector orthogonal to $\hat{u}$.

To realize the $\mathfrak{so}(1,3)$ algebra at semi-classical limit, we adopt the Kirillov--Kostant--Souriau (KKS) form~\cite{kirillov2025lectures}, where the Lie-algebra commutator is reinterpreted as a Poisson bracket.
The explicit construction based on the KKS form proceeds as
follows.
In semi-classical limit, we have
\begin{align}\label{eq:Poisson_fmunurho}
    \{a^\mu,a^\nu\} = f^{\mu\nu}{}_{\rho}\,a^\rho,
\end{align}
where $f^{\mu\nu}{}_{\rho}$ are the structure constants of $\mathfrak{so}(1,3)$.
Naively, we have a Poisson tensor $\omega^{\mu\nu}\equiv f^{\mu\nu}{}_{\rho}\,A^\rho$ defined on the six-dimensional embedding space. 
However, $\omega^{\mu\nu}$ is degenerate due to the presence of the two Casimir functions $C_1$ and $C_2$, but becomes non-degenerate when restricted to the four-dimensional orbit $\mathcal{M}_4$ with fixed $(C_1,C_2)$.

We parameterize the induced Poisson bracket on $\mathcal{M}_4$ as
\begin{align}
    \{F(s),G(s)\}_{\mathcal{M}_4}
    = \omega^{ab}(s)\,\partial_{s^a}F(s)\,\partial_{s^b}G(s),
\end{align}
where $\omega^{ab}$ is the non-degenerate Poisson tensor, i.e. the inverse of the symplectic two-form $\omega_{ab}$ on $\mathcal{M}_4$. 
In particular, the left hand side of \eqref{eq:Poisson_fmunurho} can be understood as
\begin{equation}\label{eq:Poisson_m4}
    \{a^\mu(s),a^\nu(s)\}_{\mathcal{M}_4}
    = \omega^{ab}(s)\,\partial_{s^a}a^\mu(s)\,\partial_{s^b}a^\nu(s).
\end{equation}
This with \eqref{eq:Poisson_fmunurho} implies the condition
\begin{align}\label{eq:omega_ab_condition}
    \omega^{ab}(s)\,\partial_{s^a}a^\mu(s)\,\partial_{s^b}a^\nu(s)
    = f^{\mu\nu}{}_{\rho}\,a^\rho(s).
\end{align}
Given the explicit chart~\eqref{eq:chart}, \eqref{eq:omega_ab_condition}
determines the non-degenerate Poisson tensor $\omega^{ab}$ on $\mathcal{M}_4$.

\subsection{Test scalar and effective metric on $\mathcal{M}_4$}

Following the strategy of \cite{Steinacker:2007dq}, we extract the emergent geometry associated with the $\mathfrak{so}(1,3)$ background by introducing a test scalar matrix $\Phi$ in the adjoint representation. 
Concretely, we add
\begin{align}
    S_\Phi
    = \frac{1}{2}\,\mathrm{tr}\!\left(\delta_{\mu\nu}\,[A^\mu,\Phi]\,[A^\nu,\Phi]\right)
\end{align}
to the bosonic action \eqref{eq:bosonic_action}. 
The equation of motion for $\Phi$ is
\begin{align}\label{eq:delta_XA_Phi}
    \delta_{\mu\nu}\,[A^\mu,[A^\nu,\Phi]] = 0.
\end{align}

In the semi-classical limit, commutators are replaced by $i$ times the Poisson brackets on $\mathcal{M}_4$, which is defined in \eqref{eq:Poisson_m4}.
Also, the matrix $A^\mu$ is mapped to the embedding function $a^\mu$ of $\mathcal{M}_4$ with the semi-classical limit of \eqref{eq:casimirs}; the matrix $\Phi$ corresponds to a scalar field $\phi$ that propagates on $\mathcal{M}_4$.
Hence, the semi-classical limit of \eqref{eq:delta_XA_Phi} is
\begin{align}\label{eq:Poisson_Laplace_4dim}
    -\delta_{\mu\nu}\,
    \{a^\mu(s),\{a^\nu(s),\phi(s)\}_{\mathcal{M}_4}\}_{\mathcal{M}_4} = 0.
\end{align}
Expanding the left-hand side in local coordinates yields a second-order differential equation on $\mathcal{M}_4$ of the form
\begin{align}
    G^{ab}(s)\,\partial_a\partial_b\phi(s) + \dots = 0,
\end{align}
where the ellipsis denotes terms with at most one derivative of $\phi$. The tensor $G^{ab}(s)$ hence defines the effective inverse metric on which $\phi(s)$ propagates.

To make the structure of this emergent geometry explicit, it is convenient to use the $\mathrm{SU}(2)$ left-invariant one-forms $\sigma^i$ ($i=1,2,3$) to express $G_{ab}$ (the matrix inverse of $G^{ab}$). 
They are defined as
\begin{equation}
\begin{split}
    \sigma^1 &= \cos\chi\,d\theta + \sin\chi\,d\varphi,\\
    \sigma^2 &= \sin\chi\,d\theta - \cos\chi\,d\varphi,\\
    \sigma^3 &= d\chi + \cos\theta\,d\varphi,
\end{split}
\end{equation}
and we also define the shifted one-forms
\begin{equation}
\begin{split}
    \tilde{\sigma}^1 &= \sigma^1
      - \frac{C_2\,r}{(C_1 + r^2)\sqrt{r^4 + C_1 r^2 - C_2^2}}\,dr,\\
    \tilde{\sigma}^3 &= \sigma^3
      + \frac{C_2}{\sqrt{r^4 + C_1 r^2 - C_2^2}}\,\sigma^2.
\end{split}
\end{equation}
In the basis $(dr,\tilde{\sigma}^1,\sigma^2,\tilde{\sigma}^3)$, $G_{ab}$ takes a diagonal form,
\begin{align}\label{eq:Gab}
    G_{ab}
    = \mathrm{diag}\!\left(
        \frac{r^2}{A(r)},\,
        1,\,
        \frac{B(r)}{C(r)},\,
        \frac{A(r)}{B(r)\,C(r)}
      \right),
\end{align}
where 
\begin{align}
    A(r) &= r^4 + C_1 r^2 - C_2^2,\\
    B(r) &= C_1 + r^2,\\
    C(r) &= C_1 + 2 r^2.
\end{align}
Since the effective metric $G_{ab}$ \eqref{eq:Gab} only depends on $r$, $G_{ab}$ is manifestly $\mathrm{SU}(2)$ isometric.

\begin{figure}[!h]
    \centering
    \includegraphics[width=0.5\textwidth]{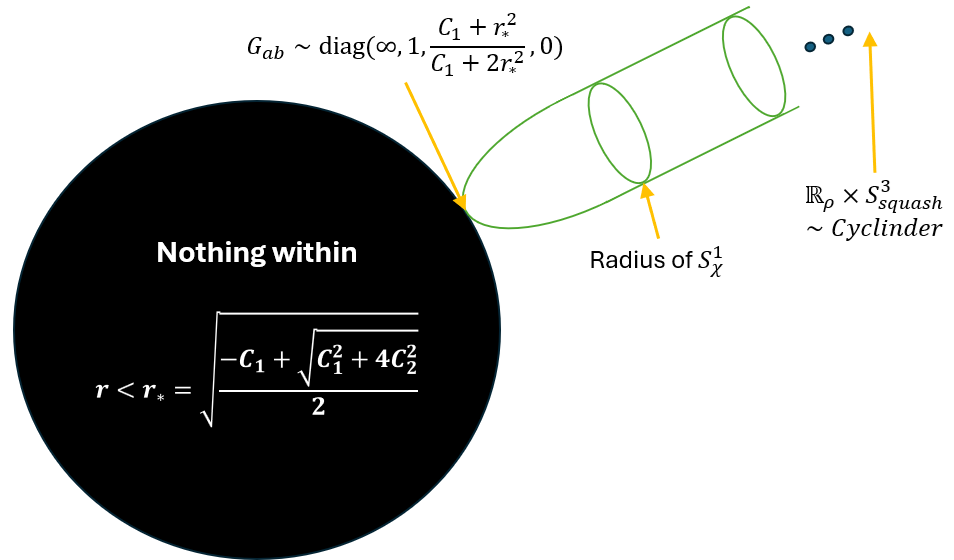}
    \caption{Geometry of effective metric~\eqref{eq:Gab}. Here, we emphasize the relation between the radius of $\chi$ direction and $r$, so each point in the plot is an $S_{\mathrm{squash}}^2$ parametrized by $\theta, \varphi$. Note that one can easily check that the curvature is finite everywhere.}
    \label{fig:effective_metric}
\end{figure}

There are two asymptotic regions that are useful to understand the geometry.
First, at $r\rightarrow \infty$ limit, the metric becomes
\begin{align}
    \lim_{r\rightarrow\infty}G_{ab}=\mathrm{diag}\left(\frac{1}{r^2}, 1, \frac{1}{2}, \frac{1}{2}\right),
\end{align}
which, upon taking $\rho=\log{r}$, describes $\mathbb{R}_\rho\times S^3_{\mathrm{squash}}$; hence, it is asymptotically cylindrical.
Second, as we approach the boundary $r=r_*$, the radius of $d\chi^2$ part of the metric shrinks to zero, which makes the metric singular.
More specifically, the metric becomes
\begin{align}
    \lim_{r\rightarrow r_*}G_{ab}=\mathrm{diag}\left(\infty, 1, \frac{C_1+r_*{}^2}{C_1+2r_*{}^2}, 0\right)
\end{align}
with finite curvatures.
We summarize these features in Fig.~\ref{fig:effective_metric}.

\section{Conclusion and outlook}\label{sec:conclusion_and_outlook}

In this letter, we have discussed an unexpected and interesting solution of the Euclidean IKKT matrix model, namely the $\mathfrak{so}(1,3)$ solution. 
It forms a completely new class of solutions in the theory, since it satisfies the equation of motion \eqref{eq:ikkt_eom} only after the summation is performed, see Sec.~\ref{sec:classification}. 

To read off the effective metric of $\mathfrak{so}(1,3)$ background, we introduce a test matrix as in~\cite{Steinacker:2007dq} to identify how it propagates at semi-classical limit.
The resulting geometry is four-dimensional, since the Lie algebra $\mathfrak{so}(1,3)$ has six generators and two Casimir constraints. 
Furthermore, such geometry, described in Fig.~\ref{fig:effective_metric}, is $\mathrm{SU}(2)$ isometric, which is closely related to the Taub--NUT/Bolt geometry and thus could be related to black hole physics.

In the following, we discuss several unresolved issues in the present study, as well as possible directions for future research, from various perspectives.

\paragraph*{(i) Effective metric}
We used the KKS form to read off the effective metric as in~\cite{Steinacker:2007dq}, but this is not the only possible approach. For example, an alternative prescription has been proposed in which the metric is defined based on the block structure of matrices~\cite{Brahma:2022dsd}. It is not yet clear how such differences in the extraction procedure affect the resulting effective metric, and further investigation will be required.

\paragraph*{(ii) Stability of the saddle} 
All analyses in this paper are based on tree-level calculations; however, the stability of the saddle should be discussed with quantum effects taken into account. 
According to~\cite{Blaschke:2011qu}, as long as the emergent spacetime has dimension at most four, the fermionic sector of the IKKT matrix model ensures that no one-loop divergences arise. 
Therefore, we expect that this saddle is stable at one-loop level.

\paragraph*{(iii) Numerical computations}
Beyond the one-loop analysis around the $\mathfrak{so}(1,3)$ saddle, understanding how dominant this configuration is within the full matrix integral will require summing over the contributions of the infinitely many other saddles as well. Such calculations will in practice need numerical analysis, for which methods such as the generalized Lefschetz-thimble approach~\cite{Cristoforetti:2012su, Alexandru:2015sua}, the complex Langevin method~\cite{Parisi:1983mgm, Klauder:1983sp} as in~\cite{Anagnostopoulos:2022dak, Chou:2025moy} or the bootstrap method~\cite{Lin:2020mme, Kazakov:2021lel}. 

\paragraph*{(iv) Relation to holography}
For the massless IKKT matrix model, the holographic dictionary was put forward in~\cite{Ciceri:2025maa, Ciceri:2025wpb}. Since our $\mathfrak{so}(1,3)$ saddle is an exact solution of this model, it is expected that the corresponding supergravity solution can be constructed by invoking the holographic principle.

At the same time, the $\mathfrak{so}(1,3)$ generators cannot be a solution of the polarised IKKT matrix model, even with large-$N$ limit. In~\cite{Hartnoll:2024csr, Komatsu:2024ydh}, the authors calculated partition function of the polarised IKKT matrix model. Furthermore, a conjecture to define the IKKT matrix model through massless limit of the polarised counterpart is made in~\cite{Komatsu:2024ydh}, where they proved for the case of $N=2,3$. However, as we have demonstrated in this work, there can be nontrivial saddles at $N\rightarrow\infty$ limit, which may alter the calculation in their work.

\section*{Acknowledgments}  
The authors are thankful for the helpful discussions with Prof. Hikaru Kawai and Prof. Pei-Ming Ho.
H.L. is supported in part by the Ministry of Science and Technology grant 112-2112-M-002-024-MY3 and 112-2628-M-002-003-MY3. R.M. is
supported by JST SPRING, Grant Number JPMJSP2132. 

\pagenumbering{alph}

\twocolumngrid
\bibliography{refs}

\end{document}